\documentclass[10pt]{article}
\usepackage[OE]{express}
\usepackage{float}

\begin{document}
\title{Optimizing the optical imaging system by \emph{in-situ} imaging the plugged hole in the ultracold atoms}

\author{Tianyou Gao,\authormark{1,2} Dongfang Zhang,\authormark{1,*} Lingran Kong,\authormark{1,2} Ruizong Li,\authormark{1,2} and Kaijun Jiang\authormark{1,3,$\dagger$}}

\address{\authormark{1}State Key Laboratory of Magnetic Resonance and Atomic and Molecular Physics, Wuhan Institute of Physics and Mathematics,
Chinese Academy of Sciences, Wuhan, 430071, China\\
\authormark{2}School of Physics, University of Chinese Academy of Sciences, Beijing 100049, China\\
\authormark{3}Center for Cold Atom Physics, Chinese Academy of Sciences, Wuhan, 430071, China
}

\email{\authormark{*}zdf0116@wipm.ac.cn} 
\email{\authormark{$\dagger$}kjjiang@wipm.ac.cn} 



\begin{abstract}
Optical absorption imaging has become a common technique for detecting the density distribution of ultracold atoms. The defocus effect generally produces artificial spatial structures in the obtained images, which confuses our understanding of the quantum systems. Here we experimentally demonstrate one method to optimize the optical imaging system by \emph{in-situ} imaging the plugged hole in the cold atoms. The atoms confined in a magnetic trap are cooled to tens of or several microkelvin by the radio-frequency evaporation cooling, and then are plugged using a blue-detuned laser beam, forming a hole in the center of the atomic cloud. We image the hole with a charge-coupled device (CCD) and quantitatively analyze the artificial spatial structure due to the defocus effect. Through minimizing the artificial structures by precisely adjusting the CCD position, we can optimize the imaging system with an accuracy of 0.1 mm. We also demonstrate the necessity of this method in probing rubidium BEC with a time of flight (TOF) of 5 ms. Compared to other methods in focusing the imaging system, the proposal demonstrated in this paper is simple and efficient, particularly for experimentally extracting large-scale parameters like atomic density, atomic number and the size of the atomic cloud.
\end{abstract}



\section{Introduction}

Since experimentally realizing Bose-Einstein condensate (BEC) in alkali metals \cite{Wieman1995Science, Ketterle1995PRL, Hulet1995PRL}, ultracold atoms have become the tabletop to study the divergent quantum effects in dilute gases, such as atomic interferometer \cite{Kaservich2015Nature}, high precision atomic clock \cite{Ye2014Nature, Ludlow2013Science}, precise spectroscopy \cite{Zelevinsky2015PRL}, quantum simulation \cite{Bloch2008RMP}, ultracold chemistry \cite{Jin2012ChemRev, Jin2011NaturePhysics}, equation of state \cite{Ueda2010Science, Salomon2010Nature}, polaron behavior \cite{Zwierlein2009PRL, Salomon2009PRL}, spin-orbit coupling \cite{Spielman2011Nature, Jiang2012PRA}, and p-wave or higher partial wave interaction \cite{Esslinger2005PRL, Jiang2014PRL}. In these studies, optical absorption imaging has become a common technique for detecting the density distribution of ultracold atoms \cite{Ketterle1999Arxiv}, providing database for extracting other physical parameters, such as size, number, density, temperature, pressure, chemical potential, entropy and so on. However, the defocus effect generally produces artificial spatial structures in the obtained images, which would mislead our understanding of the quantum systems. For example, optical diffraction patterns due to the defocus effect in ultracold atoms have been confused with the self-interference between different parts of the condensate with a long-range phase coherence \cite{Chin2011NewJ, Shin2012PRL, Langen2013PRL, Shin2013PRL}. So calibrating the imaging system and eliminating the artificial diffraction fringes are required before starting experiments on ultracold atoms.

Different methods have been used to improve the absorption imaging system previously. First, people can excite vortices in a two-dimensional (2D) gas and optimize the imaging system by minimizing the average size of the vortices \cite{Shin2014PRA, Shin2014JKPS}. This method is valid only for expanding condensate with a falling time larger than 10 ms, in which the vortex core visibility is reasonable high. It is also technically challenging due to the complexity which incorporates several advanced techniques including producing a 2D gas and exciting stable vortices. Secondly, people can optimally focus cold atom systems using the density-density correlation with a high accuracy \cite{Spielman2014RSI, Shin2014PRA}. It can works well only for freely expanding condensate with a long time of flight (TOF) and the data analysis is complicated, taking the Fourier transferring of the density distribution to get the spatial power-density distribution (PSD). Owing to the weak dependence of large-scale parameters such as peak density or width on slight defocus, such precise focusing is not necessary in many experiments. On the other hand, \emph{in-situ} imaging atoms in the trap has become an important technique in exploring quantum behaviors of ultracold quantum gas \cite{Ho2010Naturephy, Salomon2010Nature}. So optimizing the \emph{in-situ} imaging system is necessary. In this paper, we simply plug a blue-detuned laser beam with a diameter of about 20 $\mu$m into the ultracold atoms, forming a hole in the center. Then the hole is regarded as the reference during \emph{in-situ} imaging the cold atoms with a charge-coupled device (CCD). Through eliminating artificial structures created around the hole, we can optimize the imaging system with an accuracy of 0.1 mm. We also experimentally demonstrate the validity of this method by probing the density distribution of the rubidium87 BEC in focus and out of focus. Our proposal is simple and efficient particularly for experimentally extracting large-scale parameters like atomic density, atomic number and the size of the atomic cloud.

\section{Experimental setup}

\begin{figure}[htb]
\centerline{\includegraphics[width=0.93\columnwidth]{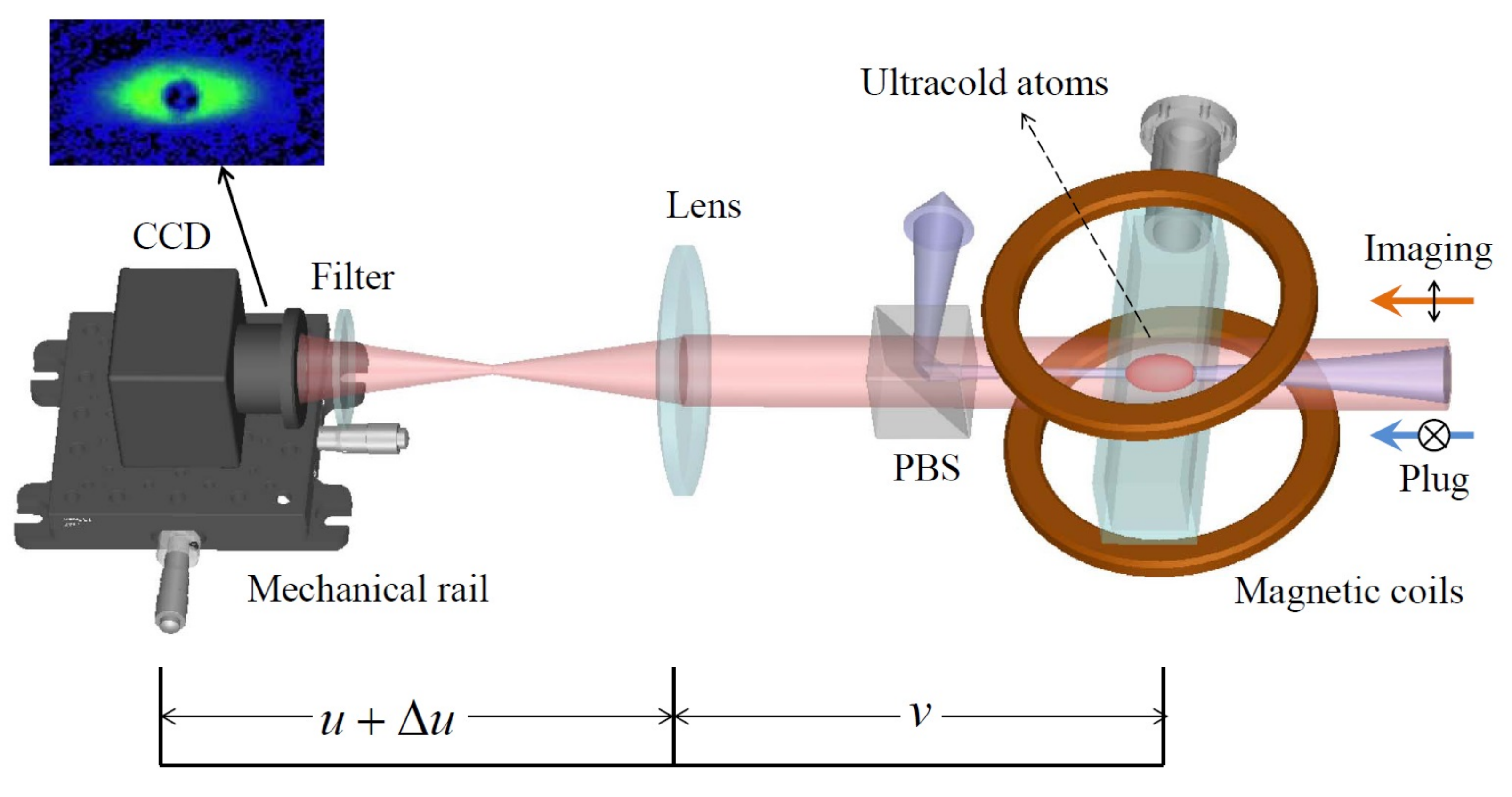}}
\caption{(Color online) Schematics of the experimental setup. CCD: charge-coupled device, PBS: polarization beam splitter. The RF signal for the evaporation cooling is not shown for the simplicity. The left-top picture shows the image of cold atoms with a plugged hole.}
\label{Fig1}
\end{figure}

Rubidium BEC can be produced in an optically plugged magnetic quadrupole trap (OPQT) as described in our previous work \cite{Jiang2016CPL}. Here we focus on how to optimize the optical imaging system using the plugged hole as a imaging reference. The experimental setup is shown in Fig.\ref{Fig1}. Ultracold rubidium atoms are confined in a magnetic quadrupole trap provided by a pair of magnetic coils. The vacuum pressure in the glass chamber is $2.0\times10^{-11}$ Torr, and the atomic lifetime in the magnetic trap is about 70 s which is long enough for the radio frequency (RF) evaporation cooling. When the radio frequency is scanned down, the atomic temperature decreases proportionally as the hot atoms are evaporated away by the RF driving spin-flip transition. A far blue-detuned laser beam with a wavelength of 760 nm and a power of 200 mW is plugged in the center of the atomic cloud. The plugged laser beam is focused into a waist of 20 $\mu$m to increase the optical barrier height. A hole forms when the atomic temperature is low enough, which is shown on the left-top side of the Fig.\ref{Fig1}. The detuning and power of the plug beam are both much smaller than those in reference \cite{Jiang2016CPL}, which greatly improves the experimental stability and simplicity. The hole position exactly overlaps the the zero point of the magnetic field, suppressing the spin-flip loss by pushing atoms away from approaching the hole. We use the standard absorption imaging method to monitor the hole in atoms. The atoms in the magnetic trap are initially in the spin state $|F=1,m_{F}=-1\rangle$, and then are optically pumped to the state $|F=2\rangle$ before being imaged.

A collimated resonant probing beam with a wavelength of 780 nm is shined on the atomic cloud. After propagating through cold atoms, the probe beam and the plug beam with orthogonal polarizations are spatially separated by a polarization beam splitter (PBS). Only the probing beam goes through a lens and is detected by a CCD. A optical filter with a bandwidth of 10 nm (Band No: Thorlabs-FBH780-10) is inserted in front of the CCD to block any light away from 780 nm, especially for the plug beam. The imaging system is in the standard 4-f configuration, where f=200 mm is the focus length of the lens. The depth of field of the imaging system is 870 $\mu$m, and the spatial resolution is 7.6 $\mu$m which is comparable to the CCD pixel size 6.8 $\mu$m. The object distance $v=2f$ equals to the image distance $u+\Delta u=2f$. $v$ and $u$ are fixed and we can precisely change $\Delta u$ by adjusting the position of the CCD which is connected on a controllable mechanical rail. When $\Delta u$ is optimized, the CCD exactly stays in the imaging plane and a perfect circular hole appears in the absorption image. If $\Delta u$ deviates from the optimized value, artificial structures will appear around the hole due to the defocus effect. Then we can optimally focus the imaging system by minimizing the artificial structures.

\section{Results and discussion}
\begin{figure}[htb]
\centerline{\includegraphics[width=0.93\columnwidth]{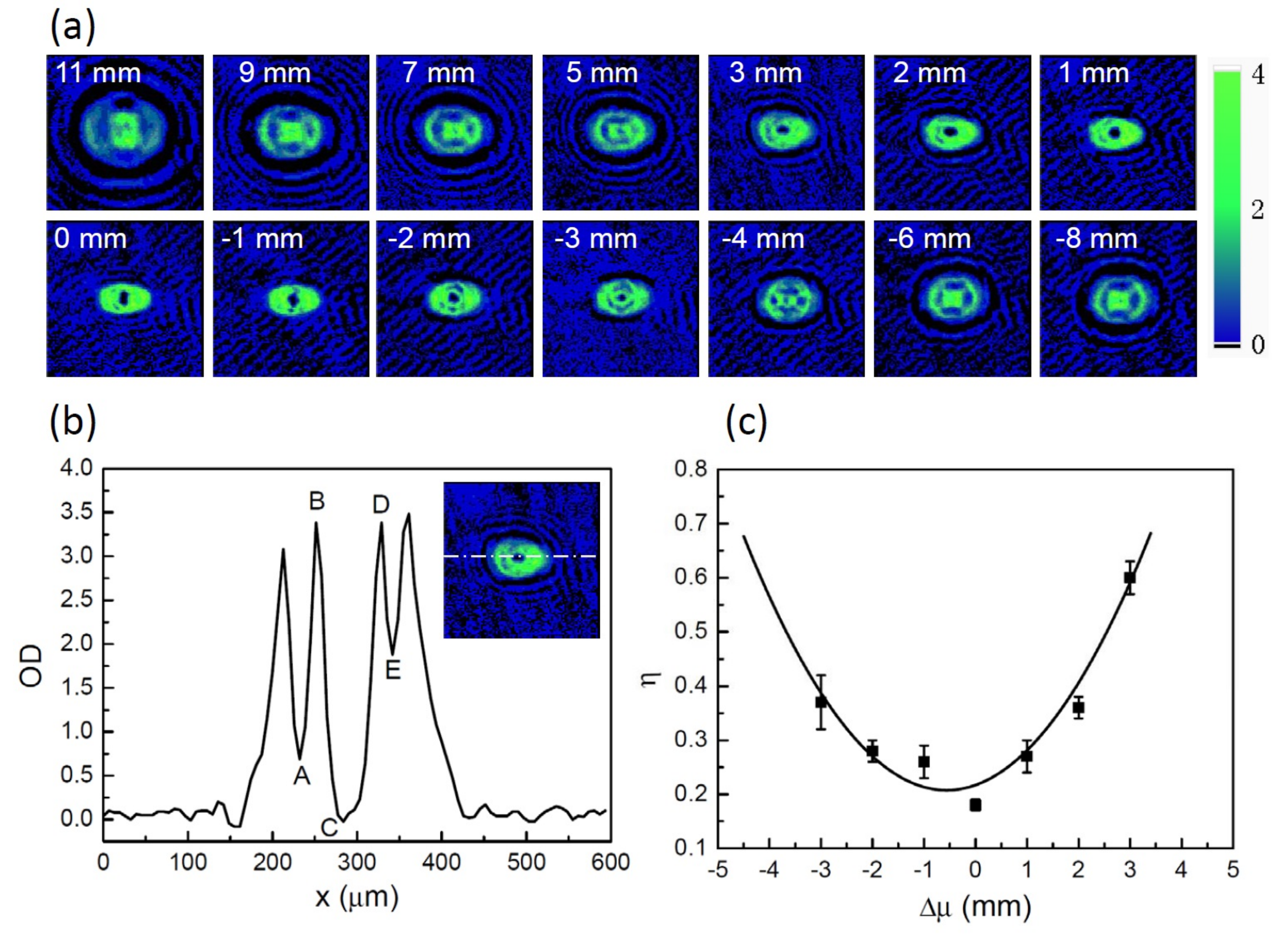}}
\caption{(Color online) Optimizing the imaging system when the RF for evaporation cooling is 1.8 MHz and the atomic temperature is 12.2 $\mu$K. (a) Absorption images for 14 different values of $\Delta u$. The color bar denotes the column optical density (OD). (b) OD distribution along the center of the atomic cloud indicated by the dot-dashed white line on the inset picture. A, B, C, D and E denote the points used to define the imaging quality factor $\eta$. The inset picture is the absorption imaging for $\Delta u$ =3 mm. (c) The imaging quality factor $\eta$ as a function of $\Delta u$. The error bars denote the standard deviation (SD) of the experimental data. The solid line is the numerical fitting with a parabolic function.}
\label{Fig2}
\end{figure}

During optimizing the absorption imaging system, $u$ and $v$ are first roughly adjusted with $u \sim 2f$ and $v \sim 2f$. When the RF for evaporation cooling is 1.8 MHz, the atomic temperature is about 12.2 $\mu$K. Fig.\ref{Fig2} (a) shows 14 absorption images with different values of $\Delta u$. One specific $\Delta u$ which is close to the optimal one is set as 0 mm. When $\Delta u$ is big like $\Delta u =$ 11 mm, 9 mm, 7 mm, 5 mm, -4 mm, -6 mm or -8 mm, the density distribution in the center area of the atomic cloud deviates strongly away from a hole, indicating that the CCD is severely away from the imaging plane. This phenomenon is similar to that in reference \cite{Shin2014JKPS} where the core of the vortex is filled up due to the strongly defocusing effect. When $\Delta u $ is small like $\Delta u =$ 3 mm, 2 mm, 1 mm, 0 mm, -1 mm, -2 mm or -3 mm, a hole clearly exists in the absorption image, but artificial interference fringes appear around the hole more or less. In order to quantitatively analyze the imaging quality, we plot the column optical density (OD) through the center of the atomic cloud (Fig.\ref{Fig2} (b)) and define the imaging quality factor as

 \begin{equation}
   \eta = \frac{(OD_B-OD_A)+(OD_D-OD_E)}{OD_B+OD_D-2OD_C}. \label{number}
\end{equation}

When CCD approaches the image plane, two dips denoted by A and E become smaller. Small $\eta$ indicates weak defocus effect and \emph{vise versa}. We plot $\eta$ in Fig.\ref{Fig2} (c) for small values of $\Delta u$. Noting that the defocusing effect should be symmetric for small defocusing distances, we use a parabolic function $\eta=a+b(\Delta u+\Delta u_0)^2$ to fit the imaging quality. $a$ and $b$ are the arbitrary fitting parameters. The numerical fitting gives $\Delta u_0 =$ 0.6(1) mm, which means that the optimal value of $\Delta u$ is -0.6 mm and the positioning precision is 0.1 mm.

\begin{figure}[htb]
\centerline{\includegraphics[width=0.83\columnwidth]{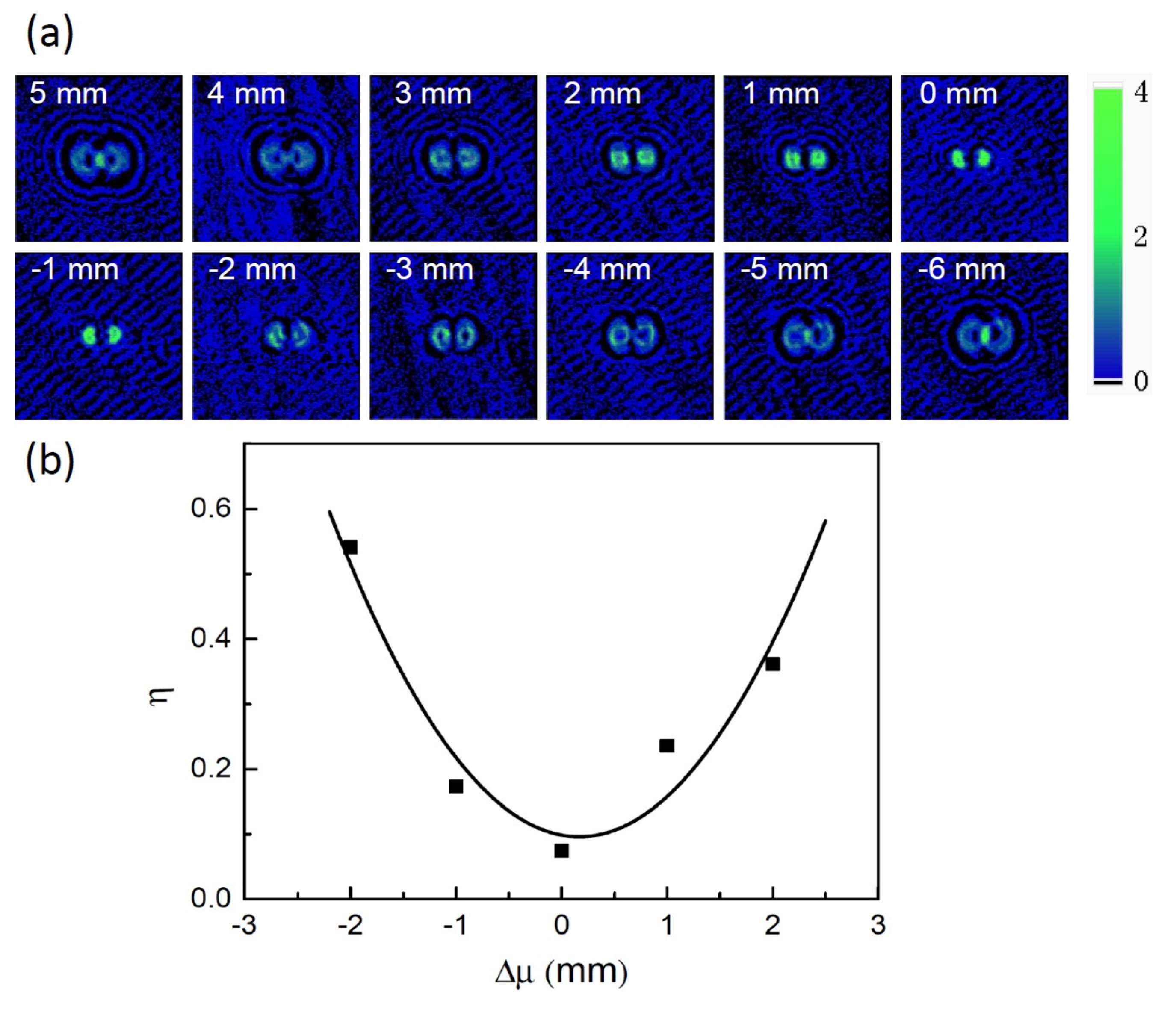}}
\caption{(Color online) Optimizing the imaging system when the RF for evaporation cooling is 1.0 MHz and the atomic temperature is 1.7 $\mu$K. (a) Absorption images for 12 different values of $\Delta u$, The color bar denotes the column optical density. (b) The imaging quality factor $\eta$ as a function of $\Delta u$. The solid line indicates the numerical fitting with a parabolic function.}
\label{Fig3}
\end{figure}

We further decrease the atomic temperature and optimize the imaging system using the same process. In Fig.\ref{Fig3} (a) we take 12 absorption images when RF for evaporation cooling is 1.0 MHz and the atomic temperature is 1.7 $\mu$K. The plug beam splits the atomic cloud into two parts. When $\Delta u$ is big like $\Delta u =$ 5 mm, 4 mm, 3 mm, -3 mm, -4 mm, -5 mm or -6 mm, two artificial holes exist in the two separated atomic clouds. Under these conditions the CCD position is strongly away from the optimal one. For small values like $\Delta u =$ 2 mm, 1 mm, 0 mm, -1 mm, or -2 mm, we plot the imaging quality factor $\eta$ in Fig.\ref{Fig3} (b). The parabolic fitting gives $\Delta u_0 =$ -0.2(1) mm. The positioning precision is 0.1 mm which is the same with that in the case of RF=1.8 MHz. $\Delta u_0$ shifts about 0.8 mm due to the day-to-day mechanical fluctuation.

\begin{figure}[H]
\centerline{\includegraphics[width=0.93\columnwidth]{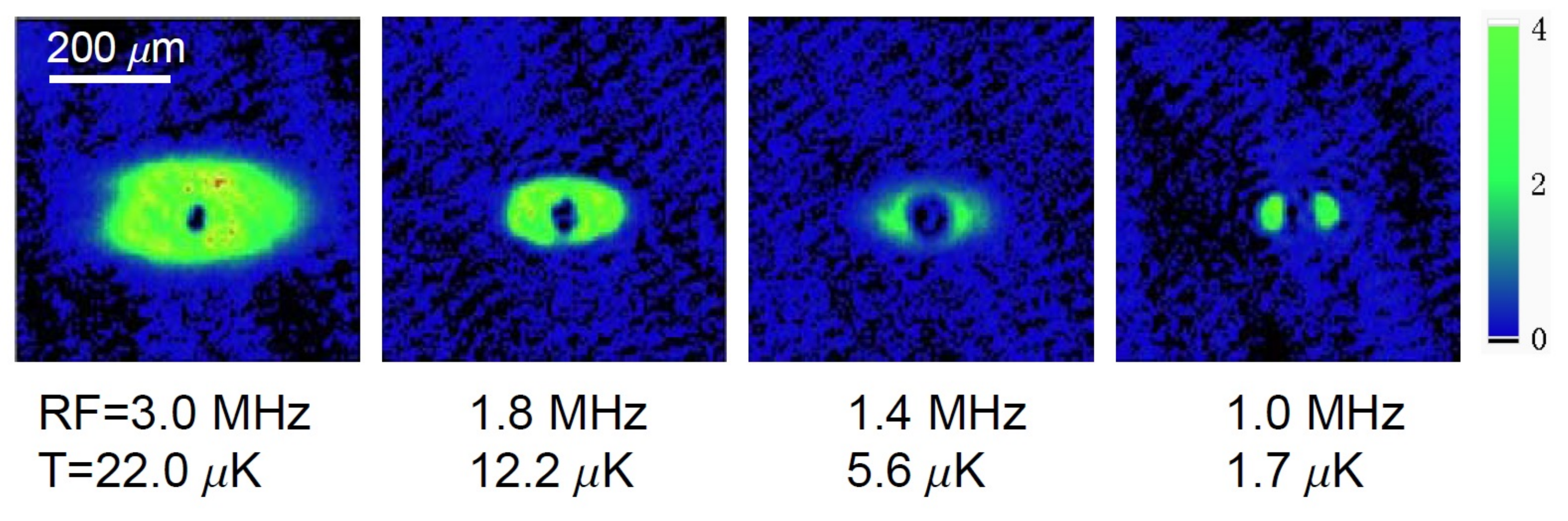}}
\caption{(Color online) Evolution of the plugged hole during the radio frequency evaporation cooling when the CCD position is optimized. The corresponding atomic temperatures are also shown. The color bar denotes the column optical density.}
\label{Fig4}
\end{figure}

In Fig.\ref{Fig4} we summarize the evolution of the hole during the period of deceasing the atomic temperature. The artificial structures surrounding the hole are negligible if the CCD position is optimized. When the RF for evaporation cooling is scanned down from 3.0 MHz to 1.4 MHz and the atomic temperature decreases from 22.0 $\mu$K to 5.6 $\mu$K, a hole clearly exists in the atomic cloud and becomes bigger with temperature. When the RF is 1.0 MHz and the atomic temperature is 1.7 $\mu$K, the plug beam splits the atomic cloud into two parts, which is the same with Fig. \ref{Fig3} (a). So probing atoms with the optimized imaging system can correctly show the dynamic process of atoms during evaporation cooling.

\begin{figure}[H]
\centerline{\includegraphics[width=0.93\columnwidth]{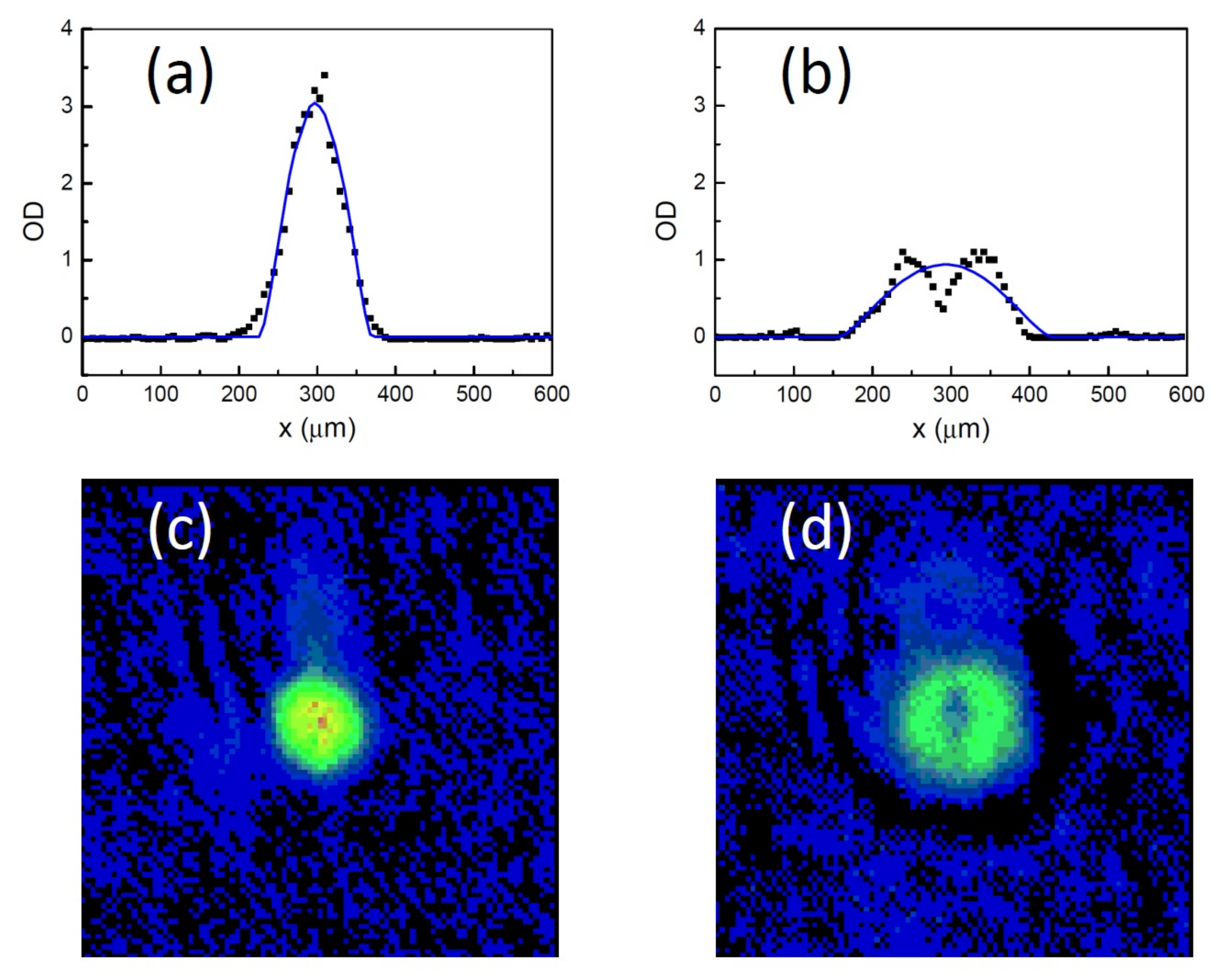}}
\caption{(Color online) Absorption images of BEC with a TOF of 5 ms. (c) is the absorption image when the CCD position is optimized and (a) is the corresponding optical density (OD) distribution through the center of the atomic cloud. Black dots denote the experimental data and the blue solid line is the numerical fitting with a Thormas-Fermi function. (b) and (d) are for the case when CCD is 3.0 mm away from the optimal position.}
\label{Fig5}
\end{figure}

In Fig.\ref{Fig4} the atomic temperature is above the BEC threshold and the atomic cloud is big. In these cases, we analyze the hole structure with the \emph{in-situ} imaging method to optimize the optical imaging system. Next we will demonstrate the necessity of this method for imaging BEC. The atomic cloud of BEC is too small that the \emph{in-situ} imaging method is no longer valid due to the limited spatial resolution. In order to analyze the density distribution of BEC, time-of-flight (TOF) probing is used as a common method, in which ultracold atoms expand freely after the trapping potential being switching off. The gravity force during the TOF has negligible effect on the imaging process because the probe light propagates horizontally in our work \cite{Shin2014PRA, Shin2014JKPS}. We first calibrate and optimize the imaging system by \emph{in-situ} detecting cold atoms confined in a magnetic trap with a plugged hole as mentioned above. Then the cold atoms are transferred into an optical dipole trap which is comprised of far red-detuned laser beams with a wavelength of 1064 nm. We produce pure rubidium BEC with a temperature of about 100 nK in the optical dipole trap \cite{Ketterle1998PRL} and there is no hole in the atomic cloud. The density distribution of BEC should be a Thormas-Fermi function, which obeys the nonlinear Gross-Pitaevskii equation \cite{Stringari1999RMP}. Fig. \ref{Fig5} shows the optical density of BEC with a TOF of 5 ms. When the CCD position is optimized, there is no artificial structure in the image and the experimental data can be well fitted with a Thormas-Fermi function (Fig. \ref{Fig5} (a) and (c)). The atomic number is $8.1\times 10^4$ and the peak density after TOF is $1.3\times 10^{11}$. The size of the atomic cloud is 68 $\mu m \times$73 $\mu m$. On the other hand, if the CCD position is 3.0 mm away from the optimal value, a hole appears in the atomic cloud due to the defocus effect and the experimental data couldn't be fitted with the Thormas-Fermi distribution (Fig. \ref{Fig5} (b) and (d)). In this case, the atomic number is $7.6\times 10^4$ and the peak density is $2.2\times 10^{10}$. The fitted size of the atomic cloud is 126 $\mu m \times$131 $\mu m$. All the physical parameters strongly deviate from the true values. The discrepancy becomes smaller when the CCD being closer to the optimal position. We find that this discrepancy is negligibly small if the distance of the CCD to the optimized position is within 1.0 mm. That means that our positioning precision including the mechanical fluctuation is high enough to probe BEC.

\section{Conclusion}
 We experimentally demonstrate one method to optimize the optical imaging system by \emph{in-situ} imaging the plugged hole in cold atoms. The atoms confined in a magnetic trap are cooled to tens of or several microkelvin by the RF evaporation cooling, and then are plugged using a blue-detuned laser beam, forming a hole in the center of the atomic cloud. We take the hole as the reference when \emph{in-situ} imaging the cold atoms and quantitatively analyze the artificial spatial structure due to the defocus effect. Through minimizing artificial structures by precisely adjusting the CCD position, we can optimize the imaging system with an accuracy of 0.1 mm. Here we show how to optimize the imaging system for atomic temperatures T=12.2 $\mu$K and T=1.7 $\mu$K. For other atomic temperatures under the condition that a hole exists, the optimizing process is the same. The optimal CCD position shifts about 0.8 mm due to the day-to-day mechanical fluctuation and less than 0.3 mm in one day. We optimize the imaging system by probing atoms with a temperature above BEC, and demonstrate its necessity in probing rubidium BEC with a TOF of 5 ms. If the CCD position is 3.0 mm away from the optimal value, large-scale parameters like peak density, atomic number and the size of the atomic cloud, will strongly deviate from the true values. Optimizing the optical imaging system as demonstrated in this paper is valid for study ultracold bosonic and fermionic gases, quantum gases in the optical lattice and ultracold atoms in low dimensions.

\section*{Funding}
This work is supported by National Key Research and Development Program of China (2016YFA0301503), and NSFC (11674358, 11434015, 91336106).

\section*{Acknowledgments}
The authors thank Kai Li and Liuyang Chen for helpful discussion about the experimental setup.

\end{document}